\documentclass{wiley-article}
\usepackage[super]{natbib}
\usepackage{hyperref}
\usepackage{siunitx}

\papertype{Preprint}

\title{Multi-modal segmentation of 3D brain scans using neural networks}

\author[1]{Jonathan Zopes}
\author[1]{Moritz Platscher}
\author[1]{Silvio Paganucci}
\author[1]{Christian Federau}

\affil[1]{Institute for Biomedical Engineering, University and ETH Zurich , Zurich, Switzerland}

\corraddress{}
\corremail{zopes@biomed.ee.ethz.ch}

\fundinginfo{This work was supported by a SPARK award of the Swiss National Science Foundation.}

\wordcount{2871}

\wordcountabstract{220}

\submittedto{Magnetic resonance imaging in medicine}

\runningauthor{Zopes et al.}

\begin{document}

\begin{frontmatter}

\maketitle
\begin{abstract}
\textbf{Purpose:} To implement a brain segmentation pipeline based on convolutional neural networks, which rapidly segments 3D volumes into 27 anatomical structures. To provide an extensive, comparative study of segmentation performance on various contrasts of magnetic resonance imaging (MRI) and computed tomography (CT) scans.\\ \textbf{Methods:} Deep convolutional neural networks are trained to segment 3D MRI (MPRAGE, DWI, FLAIR) and CT scans. A large database of in total 851 MRI/CT scans is used for neural network training. Training labels are obtained on the MPRAGE contrast and coregistered to the other imaging modalities.  The segmentation quality is quantified using the Dice metric for a total of 27 anatomical structures. Dropout sampling is implemented to identify corrupted input scans or low-quality segmentations. Full segmentation of 3D volumes with more than 2 million voxels is obtained in less than $1\,\mathrm{s}$ of processing time on a graphical processing unit.\\
\textbf{Results:} The best average Dice score is found on $T_1$-weighted MPRAGE ($85.3\pm4.6\,\%$). However, for FLAIR ($80.0\pm7.1\,\%$), DWI ($78.2\pm7.9\,\%$) and CT ($79.1\pm 7.9\,\%$), good-quality segmentation is feasible for most anatomical structures. Corrupted input volumes or low-quality segmentations can be detected using dropout sampling.\\ \textbf{Conclusion:} The flexibility and performance of deep convolutional neural networks enables the direct, real-time segmentation of FLAIR, DWI and CT scans without requiring $T_1$-weighted scans.
\end{abstract}

\end{frontmatter}

\section{Introduction}
Anatomical segmentation of magnetic resonance imaging (MRI) or computed tomography (CT) scans is important for clinical diagnostics and scientific research. In particular, quantitative volumetric measures of anatomical structures can be derived from accurate segmentation labels, which can then be used to identify and monitor the progression of diseases, such as degenerative diseases. For example, Alzheimer's disease is characterized by atrophy of the hippocampus \cite{scahill02}, Huntington disease results in athrophy of the striatum \cite{halliday98}, and frontotemporal lobar degeneration causes atrophy of the frontal and temporal lobes \cite{lu13}.

Manual brain segmentation, however, requires expert knowledge of radiologists, is extremely tedious and time consuming,  and is therefore limited to small datasets or simply not available. An alternative approach is to automatize segmentation, which sparked the development of various segmentation software packages. In brain imaging these include e.g. FreeSurfer \cite{fischl02}, BrainSuite \cite{shattuck02}, FSL \cite{jenkinson12} and ANTS \cite{avants09}. These tools apply a set of complex transformations and tresholding procedures to the input volume \cite{dale99} and are typically tailored towards $T_1$-weighted scans. As a consequence, direct segmentation of highly relevant MRI contrasts like FLAIR (fluid-attenuated inversion recovery) or  DWI (diffusion-weighted imaging) remain unsupported. The same statement is true for CT volumes.

Although the recent literature contains attempts to automatize segmentation on FLAIR \cite{korfiatis16, gibson10, duong19}, DWI \cite{cheng20, ciritsis18} or CT volumes \cite{irimia19, hu05}, a comparative study on the achievable segmentation quality on the different imaging modalities is, to the best of our knowledge, still outstanding. We attribute this to the lack of structured databases that contain several paired imaging modalities for the same patient. Further, the limited flexibility of conventional segmentation tools, that require careful fine-tuning of parameters, might be a second contributing factor.

In our work, we present a broad study on the segmentation performance achievable on $T_1$-weighted MRI, FLAIR, DWI and CT scans for a wide range of 27 anatomical classes. The analysis is based on two large databases with in total 851 MRI/CT scans and with several imaging modalities per patient. To implement a flexible segmentation pipeline, which can be quickly adapted to the different imaging modalities, we leverage the flexibility and performance of convolutional neural networks (CNNs). The recent success of CNNs in computer vision tasks \cite{krizhevsky12} provided a strong impetus for applying CNNs in brain segmentation \cite{moeskops16, mehta17}. CNNs can be rapidly adjusted to segment on a given contrast, merely by adjusting the weights of the neural network via training. This eliminates the need for additional human fine-tuning and enables us to benchmark the segmentation performance for a common network architecture (see Figure \ref{fig:fig1}). 

Further, CNN segmentation tools recently exceeded conventional processing tools in performance \cite{wachinger18} and due to their efficient implementation on graphical processing units (GPUs), achieve full segmentation of 3D volumes almost in real-time. This is orders of magnitude faster than with conventional methods \cite{roy19a}.

\section{Methods}

\subsection{Segmentation pipeline}
\begin{figure}[ht!]
    \centering
	\includegraphics[width=0.5\textwidth]{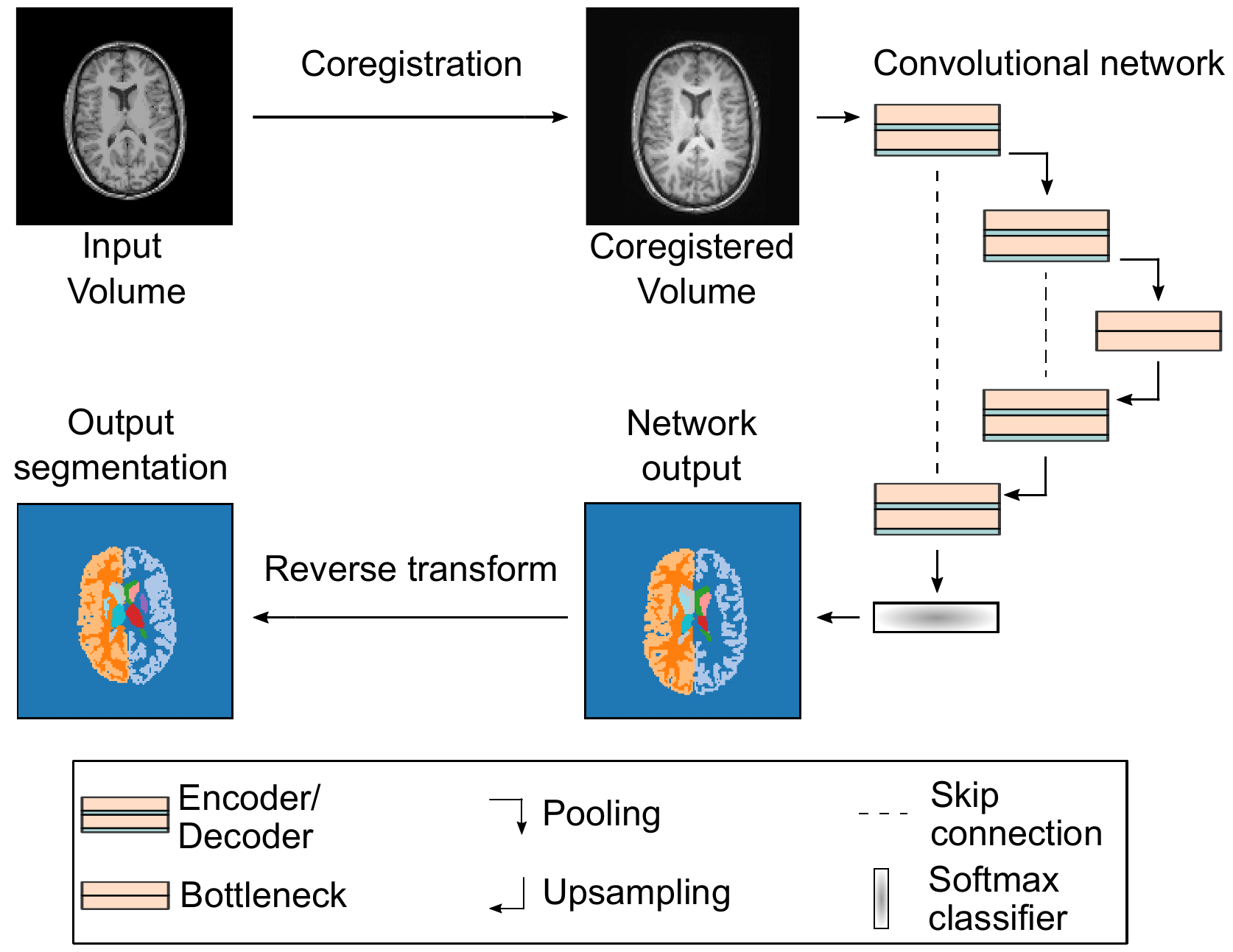}
	\caption{
\textbf{Segmentation pipeline and neural network architecture:} 3D MRI or CT input volumes are coregistered to a reference volume with an affine transformation. By proper resampling the pixel dimensions of the registered volume are adjusted to the input shape of the neural network. In addition, the pixel intensities are normalized to the interval $I = [0, 100]$. Neural networks are based on the U-Net architecture \cite{ronneberger15} with 3D convolutions in the encoder and decoder blocks.  Each encoder and decoder block contains two consecutive convolution, batch normalization and rectified linear activation operations. The encoder and decoder blocks are followed by a dropout layer (see Tab. \ref{tab:tab1} and text for details). The softmax output of the network is converted into a segmentation map with 28 labels (including background). The segmentation map is finally registered back to the input volume using the inverse affine transformation of the initial coregistration.}
	\label{fig:fig1}
\end{figure}

In Figure \ref{fig:fig1}, we show a schematic of our segmentation pipeline. The input MRI/CT volume is first coregistered to a reference volume with an affine transformation. For this task we use the registration tool elastix \cite{klein10}. The coregistered volume is resampled using spline interpolation to match the input dimensions of the segmentation CNN. The coregistration procedure increases the performance of the segmentation network and further allows for arbitrarily shaped input volumes due to resampling.

For segmentation we use a fully-convolutional neural network (F-CNN) based on the U-Net architecture \cite{ronneberger15}. A schematic of the network architecture is displayed in Figure \ref{fig:fig1} and further details on network training and parameters are discussed in the subsequent sections. The network outputs a softmax quasi-probability map $P_s(x)$ for each segmentation class $s \in \mathcal{S}$. Each individual map has the same dimension as the input image. The list of segmented classes $\mathcal{S}$ follows reference \cite{wachinger18} and comprises in total 27 structures. All segmented classes are listed in Tab. \ref{tab:tab1app}.

The softmax output $P$ of the network is converted to a hard segmentation mask $S$ using the $\mathrm{arg} \, \max$ function:
%
\begin{align}
S(x) = \mathrm{arg} \, \max_{s} \, P_s (x).
\end{align}
%
Subsequently, the hard segmentation mask $S$ is registered back to the input volume. For this purpose the initial affine coregistration transformation is inverted. After applying the inverse transformation the mask is resampled using nearest-neighbour sampling with the dimensions defined by the initial input volume.

\subsection{Neural networks and training}

As mentioned before, we use a U-Net based network architecture for segmentation. Following the findings in \cite{isensee19}, we make only minor modifications to the original implementation in \cite{ronneberger15, ozgun16}. The network consists of an encoder-decoder structure with skip connections (see Figure \ref{fig:fig1}). In each encoder and decoder block we apply two repetitions of convolutional layers, with kernel size $K = (3,\,3, \,3)$. Each convolutional layer is followed by batch normalization and non-linear activation with rectified linear units. The initial number of feature maps, after the first convolutional layer, was fixed to $F=16$ for all models and after each encoder (decoder) block the number of feature maps is doubled (halved). 

We use dropout layers after the encoders and decoders to prevent overfitting and to perform dropout sampling for uncertainty quantification (see Section \ref{sec:uncertainty}). Max pooling after each encoder block halves the feature map dimensions. Likewise, upsampling with transpose convolutions after the decoder blocks doubles the feature map dimensions and finally restores the initial dimensions at the output. 

The number of max pooling operations defines the depth $D$ of the U-Net architecture, which we fixed to $D=4$ for all trained models.  The bottleneck block restricts information flow from encoder to decoder and consists of two convolutional layers, each followed by batch normalization and rectified linear activation. In contrast to the encoder and decoder blocks, we do not use dropout layers in the bottleneck block \cite{roy19b}.

CNNs are implemented in tensorflow \cite{abadi16} and training is performed on a single GPU (Nvidia Titan RTX 24GB). Due to memory constraints the input brain volumes are limited to about 2 million voxels, which we typically distribute evenly among the imaging dimensions. The input dimensions for each network are listed in Tab. \ref{tab:tab1}. We train the network using the Adam optimizer with initial learning rates of $0.001$. During training, we apply a set of random transformations, e.g. translations, rotations or cropping, to the volumes for data augmentation. As the loss function, we use a combination of the Dice score, summed over all class labels, and the categorical cross-entropy function:
%
\begin{align}
    \mathcal{L} = - \sum_{s \in \mathcal{S}} \left( \frac{2\sum_x P_s(x) T_s(x)}{\sum_x P_s(x) + T_s(x)} - \sum_x T_s(x) \log(P_s(x)) \right).
\end{align}
%
Here, $P_s(x)$ is the softmax output of the network at voxel position $x$ and $T_s(x)$ is the ground truth at the same position. We use the categorical cross-entropy loss to alleviate convergence problems when using solely the Dice loss \cite{isensee19}. In principle, the influence of cross-entropy and Dice loss can be additionally weighted, but we found little influence on performance and therefore omit additional weighting. We train the CNNs for up to 400 epochs and abort the training process if the validation loss does not improve for 100 epochs.

\subsection{MRI/CT databases and preprocessing}

For training of the CNNs we use two large database of MRI and CT brain scans acquired on healthy patients. The acquisition parameters are listed in Table \ref{tab:tab3app}. The first database contains 530 patients of patients with normal findings (as defined by the radiological report) for which MPRAGE, FLAIR and DWI scans are available. The MPRAGE contrast was used to generate training labels using FreeSurfer \cite{fischl02}. The FreeSurfer labels were mapped to 27 segmentation classes using the mapping strategy described in \cite{roy19a}. The resulting labels are in the following considered the ground truth and subsequently coregistered to the corresponding FLAIR and DWI scans. 

After coregistration we manually checked for a proper alignment of the segmentation masks to the FLAIR or DWI volume. Out of the initial database with 530 cases, we select 164 (FLAIR) and 124 (DWI) volumes for training. We thus removed a large fraction of cases from the database. This is due to the limited fidelity of the coregistration process and because we observe that a smaller, yet higher quality database leads to better segmentation performance. For the MPRAGE contrast no further coregistration was necessary and we therefore manually selected a large fraction of 522 out of 530 volumes, with high-quality FreeSurfer segmentations, for training and testing. 

The second database contains 60 patients with normal findings for which both MPRAGE and CT brain scans are available. We again use FreeSurfer on the MPRAGE scans to obtain training labels and coregistered them subsequently to the CT volumes. By manually checking the alignment of the segmentation mask to the CT volume we selected 41 volumes for training and testing. Here, we also manually corrected minor coregistration errors to keep most of the available samples for training. For evaluation of the segmentation performance we randomly selected $10\,\%$ of the volumes in the database (see Tab. \ref{tab:tab1}).

\subsection{Segmentation performance metrics}

We use the Dice score $\mathcal{D}_s$, associated with the anatomical structure $s \in \mathcal{S}$, as the performance metric:
%
\begin{align}
\mathcal{D}_s = \frac{2 \sum_x P_s(x) T_s(x)}{\sum_x P_s(x) + T_s(x)}.
\end{align}
%
To compare the overall performance, we introduce two additional metrics: The average Dice score:
%
\begin{align}
\mathcal{D}_A = \sum_{s \in \mathcal{S}} \mathcal{D}_s,
\label{eq:average-dice}
\end{align}
%
and a volume-weighted Dice score:
%
\begin{align}
\mathcal{D}_V = \frac{1}{V}\sum_{s \in \mathcal{S}} \mathcal{V}_s \mathcal{D}_s.
\label{eq:weighted-dice}
\end{align}
%
Here, $\mathcal{V}_s$ is the volume of the structure $s$ and $V$ is the total volume of all anatomical structures $V=\sum_{s \in \mathcal{S}} \mathcal{V}_s$. We collect all resulting metrics for the different imaging modalities in Tab. \ref{tab:tab1}. The background label is not included in the average and the volume-weighted Dice score.

\subsection{Uncertainty quantification}

A common challenge for automatic segmentation tools is uncertainty quantification or quality control of the segmentation output. Low quality segmentation can occur, for example, due to corrupted input volumes, acquisition artifacts, unrecognized pathologies or in general due to input volumes outside the training distribution. The incorporation of a direct quality control method, into the segmentation process, is therefore highly desirable. 

The softmax output of neural networks, however, does not directly provide credible information on the certainty associated with the assigned labels \cite{gal16}. Instead the authors of \cite{gal16} propose to use the dropout layers of the network during prediction to make the network output stochastic. By switching some nodes off at random, we can generate a set of $N$ Monte Carlo (MC) samples ${P^1_s, ..., P^N_s}$ from the network output. The distribution of the MC samples can subsequently be used to gauge the certainty of the assigned labels. Recently, this approach has been successfully applied to brain segmentation on $T_1$-weighted MRI scans in \cite{roy19b} and we follow their methodology to equip our segmentation pipeline with a credibility metric.

To integrate dropout sampling into our segmentation pipeline we keep the dropout layers of the networks active after training. We generate $N=15$ MC segmentation samples using the, now stochastic, output of the network. The dropout rate is here fixed to $r=0.2$ for all neural networks. The final segmentation map is obtained by adding the softmax outputs of all MC samples and then applying the argmax function: 
%
\begin{align}
S(x) = \mathrm{arg} \, \max_{s} \, \sum_{i=1}^{N} P^i_s (x).
\end{align}
%
To gauge the quality of the segmentation, we use the coefficient of variation ${CV}_s$ of anatomical volumes over the MC samples. This metric was introduced in \cite{roy19b} and reads:
%
\begin{align}
{CV}_s = \frac{\sigma_s}{\mu_s}.
\end{align}
%
Here, $\sigma_s$ is the variance of the anatomical volumes between MC samples and $\mu_s$ is the mean volume. To reduce the uncertainty measure to a single quantity $CV$ we additionally average the coefficient of variation over all segmented structures:
%
\begin{align}
CV = \sum_{s \in \mathcal{S}} CV_s.
\end{align}
%

\section{Results}
\subsection{Segmentation performance}

\begin{figure*}[ht!]
    \centering
    \includegraphics[width=1\textwidth]{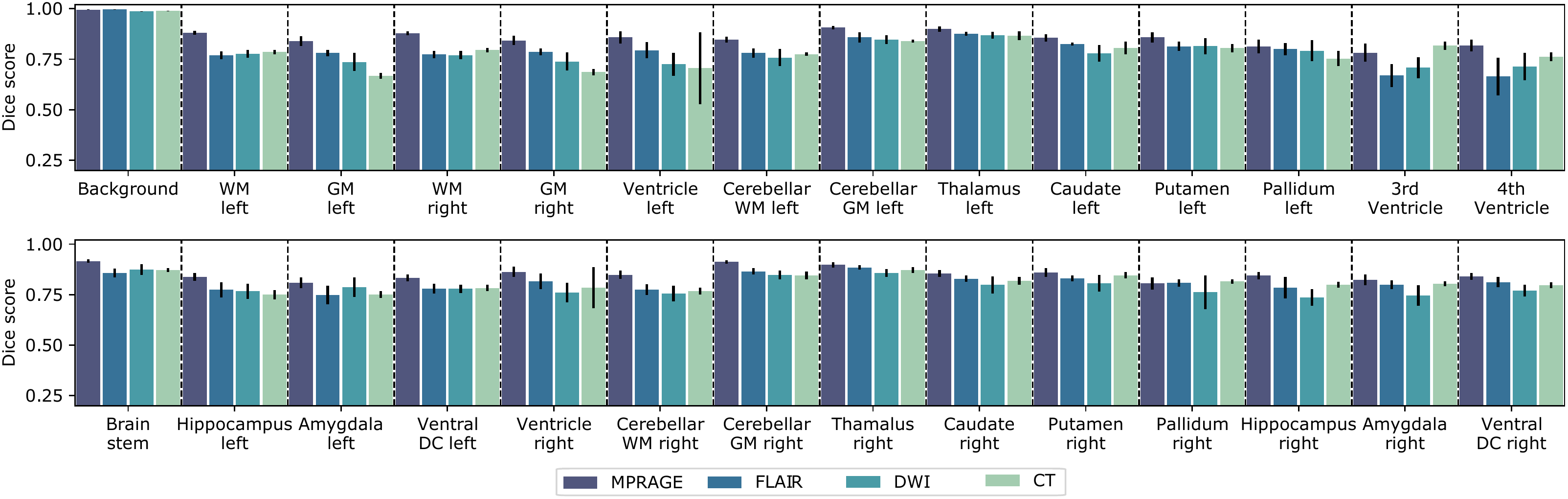}
    \caption{\textbf{Segmentation performance for different imaging modalities.} Barplot of the Dice score for all 28 segmented anatomical structures including background. For each imaging modality a separate neural network was trained and evaluated. Error bars extend from the lower to upper quartile values of the data. The Dice scores were computed from the test datasets for the corresponding imaging modality, which included 52 (MPRAGE), 16 (DWI), 12 (FLAIR) and 4 (CT) samples, respectively. All parameters of the trained CNNs are summarized in Tab. \ref{tab:tab1}.}
    \label{fig:fig2}
\end{figure*}

In Figure \ref{fig:fig2}, we compare the performance of the segmentation networks on the different imaging modalities for all 27 labelled structures and the background. The reported Dice scores represent the mean over all scans from the test set and the error bars extend from the lower to the upper quartile of values. We further collect all resulting metrics for the different imaging modalities in Tab. \ref{tab:tab1}.

We find that the best segmentation results are obtained for $T_1$-weighted, MPRAGE scans for almost all investigated anatomical structures.  This is also expressed by the best average Dice score $\mathcal{D}_A(\mathrm{MPRAGE}) = (85.3 \pm 4.6) \,\%$ and the best volume-weighted Dice score $\mathcal{D}_V(\mathrm{MPRAGE}) = (86.6 \pm 4.3) \,\%$. Second-best performance is achieved on the FLAIR contrast. Here, the average Dice score is $\mathcal{D}_A (\mathrm{FLAIR}) = (80.0 \pm 7.1)\,\%$ and the volume-weighted Dice score is $\mathcal{D}_V (\mathrm{FLAIR}) = (79.2 \pm 7.3) \,\%$. However, the difference to the performance on the DWI contrast with $\mathcal{D}_A (\mathrm{DWI}) = (78.2 \pm 7.9)\,\%$  and $\mathcal{D}_V (\mathrm{DWI}) = (77.8 \pm 8.2)\,\%$ is small.

For CT scans, we find that the segmentation performance is strongly structure-dependent: The low signal contrast between gray and white matter limits to some extent the accuracy of the  segmentation, especially of the gray matter regions. At the same time, the segmentation of structures like e.g. ventricles, the putamen or the hippocampus can be performed with high accuracy. As a consequence, we find the third-best average Dice score $\mathcal{D}_A(\mathrm{CT}) = (79.1 \pm 7.9) \,\%$ on the CT dataset, which exceeds the performance on the DWI dataset. The volume-weighted Dice score of $\mathcal{D}_V(\mathrm{CT}) = (75.8 \pm 8.2) \,\%$, however, is the lowest score in our study, due to the large volume fraction of gray and white matter.

\subsection{Example segmentations}

\begin{figure*}[ht!]
    \centering
    \includegraphics[width=1\textwidth]{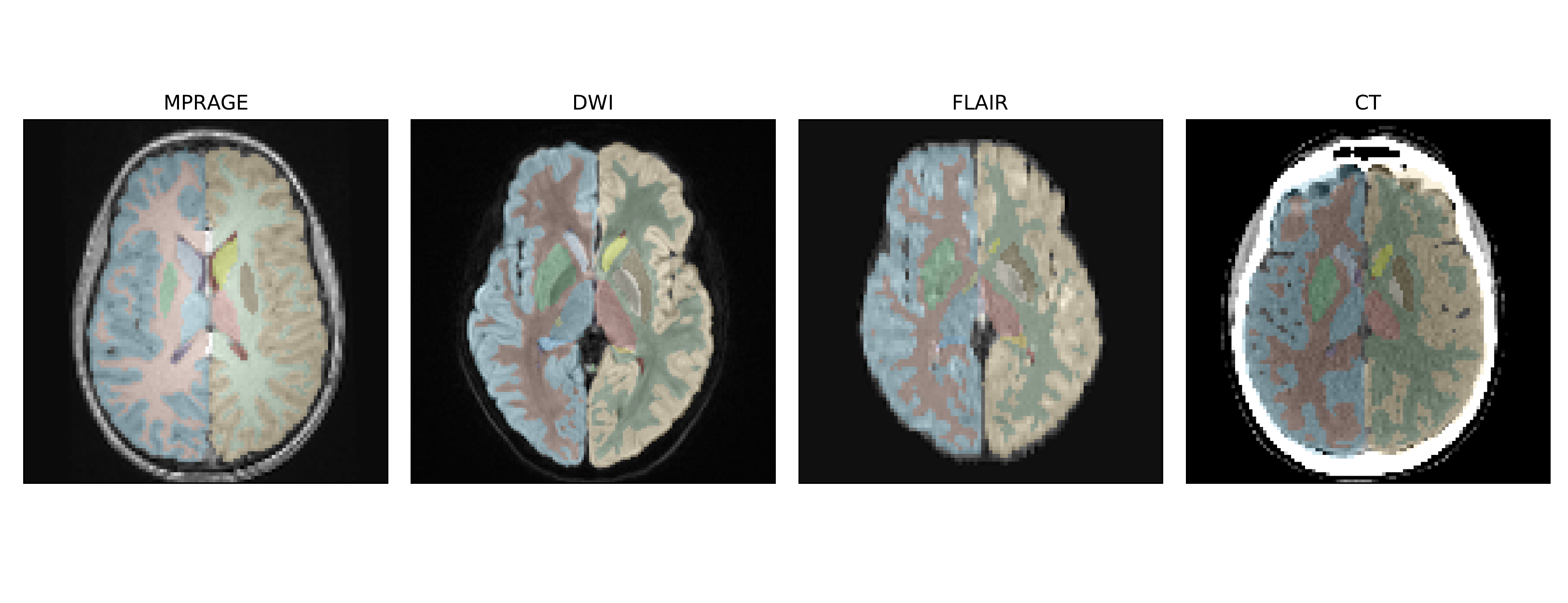}
    \caption{\textbf{Axial view at the level of the basal ganglia of brain segmentations on MPRAGE, FLAIR, DWI and CT.} The MRI were obtained from the same patient, the CT image stems from a different patient. The thalamus, the nucleus lentiformis, the nucleus caudatus and the cortical ribbon are well demarcated on all contrast. The segmentation of the cortical ribbon on CT and DWI, where the white matter - gray matter (WM-GM) contrast is low, is less detailed compared to MPRAGE, but still of good quality.}
    \label{fig:fig3}
\end{figure*}

In Figure \ref{fig:fig3}, we show exemplary segmentation maps overlaid onto each corresponding imaging modality in the axial view. For the MPRAGE, DWI and FLAIR modalities the segmentation was performed on the same patient and approximately the same slice location is displayed. Exact overlapping of slices is not possible, because the scans are not coregistered to each other. For the CT scan a separate patient was selected from the test dataset of the second database. 

The example segmentation clearly show that the gray and white matter boundaries are captured best on the $T_1$-weighted MPRAGE contrast. Here, even fine structures are properly distinguished. On the DWI and FLAIR contrast gray and white matter are segmented with lower level of detail and with lower fidelity. Due to the significantly reduced signal contrast, the gray and white matter segmentation on the CT scans displays a further reduction in performance. In terms of anatomical structures other than gray and white matter, the CT segmentation provides excellent results. This is especially the case for the ventricles, which are segmented more precisely than on the FLAIR and DWI scans.

\subsection{Uncertainty quantification}
\label{sec:uncertainty}

\begin{figure}[ht!]
    \centering
    \includegraphics[width=0.7\textwidth]{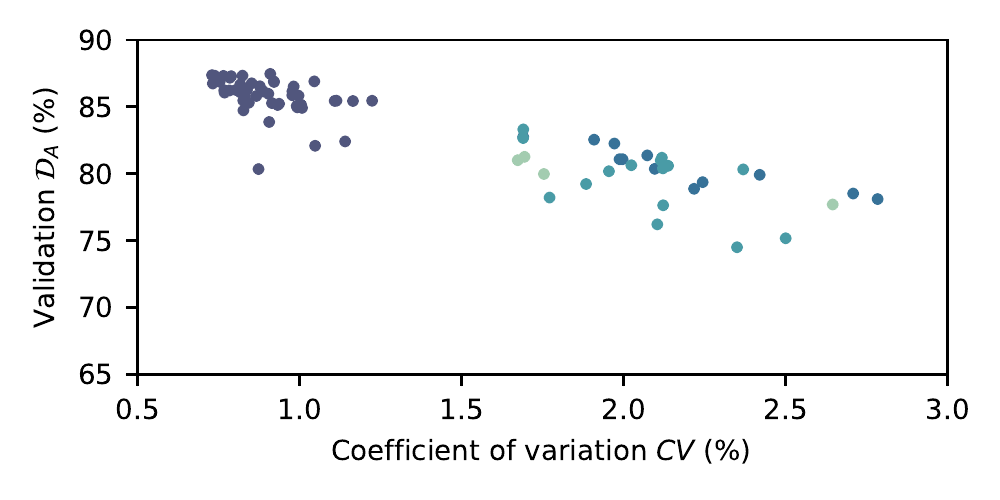}
    \caption{\textbf{Uncertainty estimation using dropout sampling.} Scatter plot of the average Dice score $\mathcal{D}_A$ versus the coefficient of variation $CV$. Each scatter point corresponds to a input volume from the test dataset. Coefficients of variation are obtained from $N=15$ MC samples. The strong correlation between $\mathcal{D}_A$ and $CV$ demonstrates that $CV$ is a good measure of segmentation quality. Pearson correlation coefficients, derived separately for each imaging modality, are summarized in Tab. \ref{tab:tab1}.}
    \label{fig:fig4}
\end{figure}

In Figure \ref{fig:fig4}, we show the relationship between uncertainty metric $CV$ and the average Dice score $\mathcal{D}_A$, derived from the ground truth labels, for volumes from the test set. Here, we combine the results for all imaging modalities. We clearly observe a strong correlation between $CV_s$ and Dice scores, which indicates that $CV$ is in fact a good metric to gauge the quality of the segmentation. The Pearson correlation coefficients are $C_{\mathrm{MPRAGE}} = -0.91$, $C_{\mathrm{FLAIR}} = -0.87$, $C_{\mathrm{DWI}} = -0.85$ and $C_{\mathrm{CT}}=-0.98$, for the corresponding imaging modalities. As a consequence, we integrate the dropout sampling as an optional processing step into our pipeline, which warns the user if the coefficient of variation $CV$ for the requested segmentation exceeds $1.0\,\%$ for MPRAGE and $2.5\%$ for FLAIR, DWI and CT contrasts, respectively.

\section{Conclusion}
We have performed a broad study on the ability of neural networks to segment brain scans acquired using MRI and CT. To our knowledge this represents the first study, which simultaneously investigates segmentation performance on a wide range of modalities and anatomical features. We find that  $T_1$-weighted images provide the best segmentation results. This finding agrees with our naive expectation, because the $T_1$-weighted MPRAGE scans provide the best gray-to-white matter contrast and the ground truth labels were generated on this contrast. Further, the largest dataset for training was available for this contrast. Nevertheless also FLAIR and DWI scans can be segmented with good performance. In case of CT scans, we observe that segmentation quality is dependent on the anatomical structure. While gray and white matter segmentation is challenging, due to low signal contrast, the performance on ventricles, putamen, pallidum and brain stem reaches or exceeds the performance achieved on the MRI contrasts. 

We further implemented dropout sampling, as introduced in \cite{roy19b}, to gauge the quality of the generated segmentation labels. We observe a strong correlation between our uncertainty metric, the coefficient of volume variation $CV$, and the quality of the segmentation derived from the ground truth labels. This is the case for all imaging modalities. Consequently, we incorporate the uncertainty metric $CV$ in our segmentation pipeline to identify faulty input volumes or low-quality segmentations.

Based on performance, flexibility and processing speed, CNNs already now represent a valuable tool for automated anatomical segmentation. In our view, the most significant obstacle to the broad applicability of segmentation CNNs is the limited generalizability to different acquisition parameters and MRI/CT scanners. To train networks that generalize very well, the the generation and distribution of large structured databases of MRI and CT scans, acquired on various scanners and imaging contrasts, is highly desirable. In addition, further research on the combination or improvement of methods, such as lifelong learning \cite{frangi18} or advanced data augmentation \cite{zhao19} is necessary. In terms of data augmentation, generative models, such as generative adverserial networks (GANs) or variational autoencoders (VAEs), could be used to generate large databases of synthetic MRI/CT scans. These databases could subsequently be used to enhance training.

\newpage

\section{Figures}

\newpage

\newpage

\newpage

\newpage
\section{Tables}
\begin{table*}[ht!]
\centering
\begin{tabular}{r|r|r|r|r|r|r}
Modality & $N_{\mathrm{train}}$ & $N_{\mathrm{test}}$  & Volume shape  & Average score $\mathcal{D}_A$ & Weighted score $\mathcal{D}_V$ & Corr $(\mathcal{D}_A, CV)$   \\ 
\hline \hline
MPRAGE           & 470 & 52                    & [128, 128, 128] & $(85.3 \pm 4.6)\,\%$ & $(86.6 \pm 4.3)\,\%$ & $-0.91$  \\
FLAIR            & 112 & 12                    & [128, 128, 128] & $(80.0 \pm 7.1)\,\%$ & $(79.2 \pm 7.3) \,\%$ & $-0.87$  \\
DWI              & 148 & 16                    & [160, 160, 32]  & $(78.2 \pm 7.9)\,\%$ & $(77.8 \pm 8.2) \,\%$ & $-0.87$  \\
CT               &  37 &  4                    & [96, 128, 128]  & $(79.1 \pm 7.9) \,\%$ & $(75.8 \pm 8.2) \,\%$ & $-0.97$   
\end{tabular}
\caption{\textbf{Parameters of training and test datasets and segmentation scores on all four imaging modalities.} Each dataset contains $N_{\mathrm{train}}$ training volumes and $N_{\mathrm{test}}$ test samples. The voxel dimension of all volumes in the training and test dataset is fixed to the reported volume shape. Average and weighted Dice scores are reported according to Eq. \ref{eq:average-dice} and \ref{eq:weighted-dice}, respectively. The Pearson correlation coefficient between average Dice score and the uncertainty metric $CV$, obtained from dropout sampling, is listed for each imaging modality.}
\label{tab:tab1}
\end{table*}

\newpage
\section{Appendix}
The appendix contains three tables with additional information on the segmented anatomical structures, on the detailed parameters of the neural networks and on the acquisition parameters of the MR scans.


\begin{table}[ht!]
\centering
\begin{tabular}{c|c}
Segmentation index & Anatomical structure          \\ \hline \hline
1                  & Cortical White Matter Left    \\
2                  & Cortical Grey Matter Left     \\
3                  & Cortical White Matter Left    \\
4                  & Cortical Grey Matter Right    \\
5                  & Lateral Ventricle Left        \\
6                  & Cerebellar White Matter Left  \\
7                  & Cerebellar Grey Matter Left   \\
8                  & Thalamus Left                 \\
9                  & Caudate Left                  \\
10                 & Putamen Left                  \\
11                 & Pallidum Left                 \\
12                 & Third Ventricle               \\
13                 & Fourth Ventricle              \\
14                 & Brainstem                     \\
15                 & Hippocampus Left              \\
16                 & Amygdala Left                 \\
17                 & Ventral DC Left               \\
18                 & Lateral Ventricle Right       \\
19                 & Cerebellar White Matter Right \\
20                 & Cerebellar Grey Matter Right  \\
21                 & Thalamus Right                \\
22                 & Caudate Right                 \\
23                 & Putamen Right                 \\
24                 & Pallidum Right                \\
25                 & Hippocampus Right             \\
26                 & Amygdala Right                \\
27                 & Ventral DC Right             
\end{tabular}
\caption{\textbf{List of all segmented anatomical structures.} The selection of anatomical structures and labelling scheme follows \cite{wachinger18}. Background labels are assigned with the index 0.}
\label{tab:tab1app}
\end{table}

\newpage
\begin{table}[ht!]
\centering
\begin{tabular}{c|c|c|c|c}
\multicolumn{1}{l|}{} & MPRAGE                                                  & FLAIR                                                   & DWI                                                  & CT                                                    \\ 
\hline \hline
Input dimension       & \begin{tabular}[c]{@{}c@{}}128\\128\\128\end{tabular} & \begin{tabular}[c]{@{}c@{}}128\\128\\128\\\end{tabular} & \begin{tabular}[c]{@{}c@{}}160\\160\\32\end{tabular} & \begin{tabular}[c]{@{}c@{}}96\\128\\128\end{tabular}  \\
\hline
Initial feature maps $F$  & 16                                                      & 16                                                      & 16                                                   & 16                                                    \\
Depth of network $D$       & 4                                                       & 4                                                       & 4                                                    & 4                                                     \\
Bottleneck layers $B$     & 2                                                       & 2                                                       & 2                                                    & 2                                                     \\
Conv. kernel size $K$    & {(}3, 3, 3)                                             & {(}3, 3, 3)                                             & {(}3, 3, 3)                                          & {(}3, 3, 3)                                          \\
Number of parameters  & 5.65M                                          			& 5.65M                                           			  & 3.25M                                           		   & 5.20M                                          		   \\
\end{tabular}
\caption{\textbf{Summary of CNN parameters.} The architecture of the CNNs follows \cite{ronneberger15}: The initial number of feature maps $F$, after the first convolutional layer, is doubled in each encoder block and halved in each of the decoder blocks. The depth $D$ of the network describes the number of down- or upsampling operations. For example, the network shown in Figure \ref{fig:fig1} represents a network of depth $D=2$. The number of bottleneck layers $B$, specifies how many convolutional layers are part of the bottleneck. In the bottleneck each convolutional layer is followed by batch normalization and rectified linear activations.}
\label{tab:tab2app}
\end{table}

\newpage
\begin{table}[ht!]
\begin{tabular}{c|c|c|c}
Parameter                              & MPRAGE               & FLAIR                & DWI                \\ \hline \hline
Repetition time $T_R$ (ms)             & $ 1938 \pm 480$      & $ 7999 \pm 711$      & $6926 \pm 781$     \\
Echo time $T_E$ (ms)                   & $ 3.25 \pm 0.54$     & $114 \pm 28$         & $73 \pm 16$        \\
Acquisition matrix (range)             & $[224-288, 184-288]$ & $[256-320, 168-320]$ & $[96-200, 96-200]$ \\
Slice thickness (mm)                   & $1.01 \pm 0.06$      & $3.88 \pm 0.53$      & $3.69 \pm 0.64$    \\
Pixel bandwidth (Hz)                   & $201 \pm 19$         & $292 \pm 45$         & $1018 \pm 211$     \\
Field strength 3T / 1.5T (\% of cases) & $95\,\% / 5\,\%$     & $100\,\% / 0\,\%$    & $92\,\% / 8\,\%$  
\end{tabular}
\caption{\textbf{Summary of MR acquisition parameters.} The repetition time ($T_R$), echo time ($T_E$), slice thickness and pixel bandwidth are reported as $\mu \pm \sigma$. Here, $\mu$ is the average value of the distribution of the parameter and $\sigma$ is the corresponding standard deviation. }
\label{tab:tab3app}
\end{table}

\end{document}